# Uniaxial strain on graphene: Raman spectroscopy study and bandgap opening

Zhen Hua Ni,[†] Ting Yu,[†]* Yun Hao Lu,[‡] Ying Ying Wang,[†] Yuan Ping Feng,[‡] Ze Xiang Shen[†]*

[†] *Division of Physics and Applied Physics, School of Physical and Mathematical Sciences, Nanyang Technological University, Singapore 637371, Singapore*
[‡] *Department of Physics, National University of Singapore, 2 Science Drive 3, Singapore 117542*

**ABSTRACT**

Graphene was deposited on a transparent and flexible substrate and tensile strain up to ~0.8% was loaded by stretching the substrate in one direction. Raman spectra of strained graphene show significant redshifts of 2D and G band (-27.8 cm$^{-1}$ and -14.2 cm$^{-1}$ per 1% strain, respectively), because of the elongation of the carbon-carbon bonds. This indicates that uniaxial strain has been successfully applied on graphene. We also proposed that by applying uniaxial strain on graphene, tunable bandgap at K point can be realized. First principle calculations predicted a bandgap opening of ~300 meV for graphene under 1% uniaxial tensile strain. The strained graphene provides an alternative way to experimentally tune the bandgap of graphene, which would be more efficient and more controllable than other methods that are used to open bandgap in graphene. Moreover, our results suggest that the flexible substrate is ready for such strain process and Raman spectroscopy can be used as an ultra-sensitive method to determine the strain.

Corresponding Authors: yuting@ntu.edu.sg; zexiang@ntu.edu.sg

Graphene consists of one flat layer of carbon atoms arranged in a honeycomb lattice, it has attracted intensive interest since experimentally discovered in 2004.[1] Due to its special properties such as the charge carriers mimicking massless relativistic Dirac fermions, the anomalous quantum Hall effect, and the ballistic transport even at room temperature,[2] graphene provides a promising future for fundamental studies and practical applications.

In order to make graphene a real technology, a special issue must be solved: creating an energy gap at $K$ and $K'$ points in the Brillouin zone. Different attempts have been made by researchers, such as patterning graphene into nanoribbon,[3] forming graphene quantum dots,[2] making use of multilayer graphene sheets and applying an external electrical field.[4] According to the carbon nanotubes (CNTs) study, the strain will dramatically change the electronic structure of CNTs.[5-7] It is shown that strain can open a bandgap in a metallic CNT and modify the bandgap in a semiconducting CNT with about 100 meV per 1% stretch.[7] Being a one-atom thick structure, it is reasonable to predict that strain, especially uniaxial strain can dramatically modify the electronic and optical properties of graphene. Moreover, since the two carbon sublattices of graphene are inequivalent under uniaxial strain, it is possible to introduce a bandgap opening on graphene due to the breaking of sublattice symmetry.[8-10] As an example, a bandgap opening is predicted for graphene growth on hexagonal boron nitride substrate due to the breaking of the equivalence of sublattice.[10]

In this study, we have successfully deposited the graphene sheets on a transparent flexible substrate: polyethylene terephthalate (PET). Uniaxial tunable tensile strain (up to ~0.8%) was applied on the single/three layer graphene by stretching the PET in one direction, as justified latter. Raman spectroscopy was used to study the strain effect on graphene. Significant red shift of Raman 2D band (-27.8 cm$^{-1}$ per 1% strain (/%)) and G band (-14.2 cm$^{-1}$/%) for single layer graphene was observed under the uniaxial tensile strain. Our first principle simulation of the band structure of single layer graphene shows a bandgap opening of 300 meV for 1% strain, which provides an alternative way to fabricate graphene-based devices.

**RESULTS AND DISCUSSION**

Figure 1 shows the Raman spectra of single and three layer graphene on PET. Raman spectra of Graphene on different substrates have been studied previously and weak dependence of Raman bands on the substrates was observed.[11, 12] In Fig. 1, the Raman fingerprint of single-layered graphene, a very sharp (~30 cm$^{-1}$) and symmetric 2D band at around 2680 cm$^{-1}$ is clearly present.[13, 14] In contrast, the 2D band of three layer graphene is much broader (~59 cm$^{-1}$) and can be fitted by multi-peaks. The change of 2D band with the increase of graphene thickness was explained by the evolution of electronic band structure of graphene[13] according to the double resonance theory. The Raman G band originates from the in-plane vibrational $E_{2g}$ phonon and locates at ~1580 cm$^{-1}$. In our work, the G band of graphene overlaps with a strong peak from PET and appears as a weak shoulder. This makes it difficult to perform a detailed and careful study, such as Raman image study. Instead, we used two lorentzian curves to fit the spectra and obtained the frequency of G band at different points of the sample, and then carried out the data analysis. The inset (a) in Fig. 1 is the optical image of the interested graphene. The optical contrast of graphene on PET is very poor comparing to the Si/SiO$_2$,[15] which makes the graphene barely seen by optical microscope. The inset Raman image (b) in Fig. 1 is constructed by the width of Raman 2D band of the chosen area, which clearly shows the single (dark region, with bandwidth about 30 cm$^{-1}$) and three layer graphene (bright region, with bandwidth about 59 cm$^{-1}$). The following strain studies were carried out on the same graphene sheets.

Figure 2a presents the 2D frequency Raman images of unstrained (a1), strained (0.18% (a2), 0.35% (a3), 0.61% (a4), and 0.78% (a5) strain), and relaxed (a6) graphene by extracting the frequency of 2D band. The Raman images were constructed by taking the peak frequencies of 2D bands of every point. The scales of all Raman images are from 2650 to 2710 cm$^{-1}$. In those images, the darker color represents lower 2D band frequency. Obviously, with the increase of strain (a2 to a5), the Raman images become darker and darker, indicating a universal red-shift of 2D band over the strained graphene. This red-shift of 2D band (and also the G band later) can be understood on the basis of the elongation of the carbon-carbon bonds, which weakens the bonds and therefore lowers their vibrational frequency. The spectra in the right-hand side of Fig 2a show the 2D band of SLG of unstrained (a1), 0.78% strained (a5), and relaxed graphene (a6). The redshift of 2D band under strain as well as the blueshift due to strain relaxation is clearly seen.

To quantify the strain coefficient, Fig. 2b shows the 2D band frequency from the same region of unstrained, strained and relaxed graphene. The mean frequencies of 2D band from the highlighted area, as shown in Fig.2b inset, were plotted as a function of strain with the standard deviation as errors. The absence of discrete jumps in the 2D band frequencies of graphene under different strain assures no slippage of graphene occurs during the stretching process. In Fig 2b, a linear dependence of 2D band frequency on strain is clearly seen, with a slope of -27.8±0.8 cm$^{-1}$/% for single layer graphene and -21.9±1.1 cm$^{-1}$/% for three layer graphene. Both of these values are comparable to that of SWNTs (-7.9 ~ -37.3 cm$^{-1}$/%).[16-18] The high strain sensitivity of graphene successfully demonstrates its potential as an ultrasensitive strain senor as has been proved on using the CNTs.[16] The linear dependence of Raman bands with strain is expected according to the phonon deformation potentials. The frequency shift of Raman band is related to the uniaxial strain and the shear strain. The shear strain has a much smaller contribution,[19] which can be ignored for simplicity. The Raman frequency is then related to the strain by:[19, 20]

$$\frac{\Delta\omega}{\omega_0} = \gamma \cdot (\varepsilon_{xx} + \varepsilon_{yy})$$

where $\gamma=1.24$ is the Güneisen parameter obtained from the experiment on CNTs.[19] $\varepsilon_{xx}$ is the uniaxial strain, while $\varepsilon_{yy} \approx -0.186\varepsilon_{xx}$ is the relative strain in the perpendicular direction according to the Poisson's ratio of graphene.[21] $\omega_0$ is the Raman band frequency. Therefore, as a rough estimation, the 2D band frequency ($\omega_0 \approx 2670$ cm$^{-1}$) dependence on uniaxial tensile strain is:

$$\frac{\Delta\omega}{\varepsilon} = -\omega_0 \cdot \gamma \cdot (1-0.186) = -26.9 cm^{-1}/\%$$

This value is very close to our experimental result of single layer graphene (-27.8cm$^{-1}$/%). The 2D frequencies of Graphene on PET are very uniform (with distribution of only 1 cm$^{-1}$) before the application of strain. However, the bigger error bars under higher strain indicate the wide distribution of 2D band frequency, reflecting the non-uniformity of local strain of single layer graphene. Once the strain is released, the 2D band shifts upwards and goes back immediately to almost the original position, indicating good strain reversibility of graphene. Such reversible and quick recovery property demonstrates the excellent elasticity of graphene, which might be critical for practical applications.

Because the G band is only a small shoulder peak, it is impossible for us to conduct the Raman imaging with the G band frequency. Raman spectra are taken from different positions (~10 positions) of the single and three layer graphene and the spectra are curve fitted to obtain the G band frequency. The frequencies of G band under different strain are shown in Fig. 3, where the bigger error bar of single layer graphene is caused by the poor signal of G band compared to the overlapping peak from PET substrate, which would introduce a large fitting error. A linear dependence of G band frequency on the strain is also clearly seen, with a slope of -14.2±0.7 cm$^{-1}$/% for single layer graphene and -12.1±0.6 cm$^{-1}$/% for three layer graphene. These values are also comparable to the results of CNTs, which is 11-17 cm$^{-1}$/%.[16]

The relative larger shifts of both G and 2D bands of single layer graphene compared to those of three layer graphene might be because strain is more effectively applied on thinner graphene sheet. The strain on graphene sheets was loaded by stretching the PET substrate, which interacts with graphene by van der Waals force. Therefore, it would be more difficult for strain to be transferred to thicker samples. One support for this interpretation is that no Raman band shift is observed on bulk graphite in the whole strain experiment, which means that the graphite is not affected by stretching the PET substrate.

Both the redshifts of 2D and G bands of graphene indicate that the graphene sheet experiences noticeable and controllable uniaxial strain by stretching the flexible substrate. The uniaxial strain in current work is different from the biaxial strain we observed in epitaxial graphene [22] and graphene after deposition of $SiO_2$ layer and annealing[23]. The uniaxial strain would affect the electronic properties of graphene much more significantly as it breaks the equivalence of sublattice of graphene.

To understand the strain dependence of electronic structure of graphene, we carried out first principle electronic band structure calculations of graphene under a uniaxial tensile strain. The schematic diagram of tensile strain application on graphene lattice is shown in Fig. 4a. A stretching in one direction results in a shrinking in another (perpendicular) direction. As an example, the band structures of unstrained and 1% strained graphene along the *Γ-K-M* are shown in Fig 4 (b) and (c), respectively. A bandgap opening of the strained graphene at *K* point of Brillouin zone is clearly seen. This bandgap opening can be attributed to the breaking of sublattice symmetry [8-10] of graphene under uniaxial strain. In general, the size of the bandgap increases almost linearly with the increase of tensile strain, as shown in Fig. 5, and for a strain of 1%, the bandgap reaches about 300 meV. The density of states of the unstrained and 1% strained graphene are also shown in the insets of Fig.5, where the bandgap in the strained graphene is clearly seen. The bandgap corresponding to the highest strain obtained in our experiment (0.78%) is indicated by the red dot and it is about 250 meV. The bandgap variation with uniaxial strain suggests that by applying a uniaxial strain, a tunable bandgap of single layer graphene can be achieved. This uniaxial strain on graphene can be easily realized by stretching the flexible substrate. The bandgap opening by uniaxial strain would be more efficient compared to other methods, such as electric field tuning on bilayer,[4] or molecule adsorption.[9] It is also easier to be realized than fabrication of graphene nanoribbons [3] and the gap is more controllable than those in epitaxial graphene[8]. The uniaxial strained graphene provides an alternative way to fabricate graphene based-devices. In addition to the flexible substrate, graphene can also be deposited on piezocrystal so that the strain can be easily and precisely controlled. Further experiments will be carried out to study the transport properties of uniaxial strained graphene and correlate with the simulated band structure.

**CONCLUSION**

In summary, we have deposited single and three layer graphene on a transparent and flexible substrate, PET. By stretching the PET, uniaxial strain of ~0.8% can be applied on graphene. Raman spectroscopy studies of strained graphene show significant redshift of G and 2D band (-14.2 cm$^{-1}$/% and -27.8 cm$^{-1}$/% respectively for single layer graphene). Finally, results of our first principle calculations suggest that a bandgap opening of ~300 meV can be achieved by applying 1% uniaxial strain on graphene. The strained graphene provides an alternative way to experimentally tune the band gap of single layer graphene, which is more efficient and more controllable than the other techniques that are used to open a bandgap in graphene.

**EXPERIMENTAL AND COMPUTATIONAL SECTION**

The graphene sample was prepared by mechanical cleavage [1, 15] from a highly ordered pyrolytic graphite (HOPG, Structure Probe, Inc./ SPI Supplies) and transferred onto a PET film. Tensile strain on graphene sheets was loaded by stretching the PET film in one direction. The amount of strain is determined by dividing the extra length with the unstrained length. Because the graphene is ultra-thin, the van der Waals force between substrate and graphene would be strong enough to exert the strain (~1%) on graphene.[24] The Raman spectra were carried out with a WITEC CRM200 Raman system. The excitation source is 532 nm laser (2.33 eV) with a laser power below 0.1 mW on the sample to avoid laser induced local heating.[25] A 100x objective lens with Numerical aperture (NA) of 0.95 was used in the Raman experiments, and the spot sizes of 532 nm laser was estimated to be 500 nm. For the Raman image, the sample was placed on an *x-y* piezostage and scanned under the illumination of laser. The Raman spectra from every spot of the sample were recorded. The stage movement and data acquisition were controlled using ScanCtrl Spectroscopy Plus software from WITec GmbH, Germany. Data analysis was done by using WITec Project software. After the application of strain, the sample was located again and Raman imaging was carried out on the same area of the sample. First principle calculations were performed using the VASP code. [26, 27] The projector augmented wave potentials were used for electron-ion interactions while the local spin density approximation (LSDA) was used for exchange-correlation function. For the Brillouin-zone integrations, we used a 45 x 45 x 1 grid of Monkhorst-Pack special points together with a Gaussian smearing of 0.1 eV for the one-electron eigenvalues. The plane wave basis set was restricted by a cutoff energy of 400 eV. The lattice constant of unstrained graphene structure was set to 0.242nm which is the lattice parameter determined from our first principle calculations within LSDA. The strained graphene was obtained by applying extension on x-axes(y-axes) with a fixed value while the value of y-axes(x-axes) was tuned as the system reaches its lowest total energy. In structure relaxation, atoms were fully relaxed until the interatomic forces are less than 0.1 eV/nm.

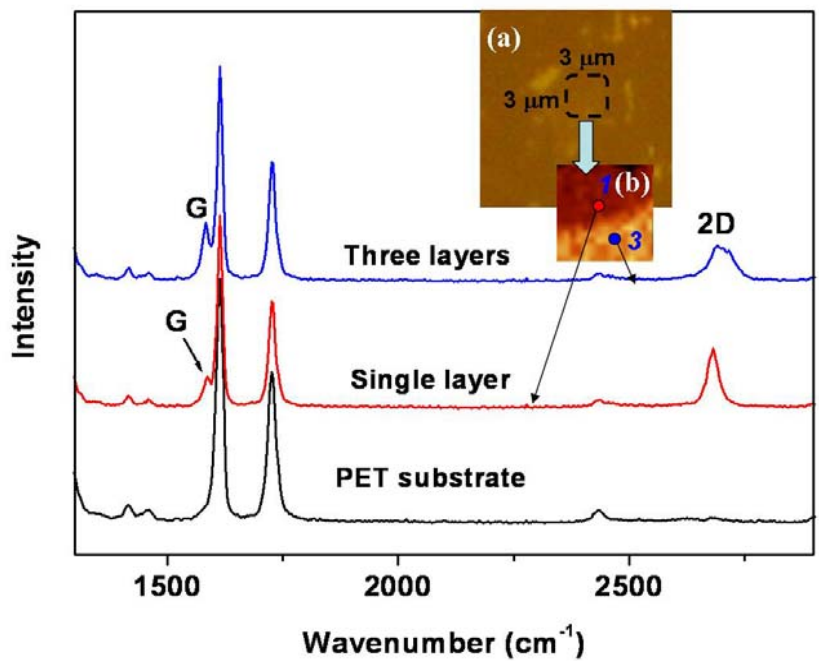

Figure 1. Raman spectra of single and three layer graphene as well as the spectrum of PET substrate. Inset (a) is the optical image of the graphene on PET substrate, with the chosen area for Raman image study. Inset (b) is the 2D width Raman image of the chosen area. The dark region with peak width of ~30 cm$^{-1}$ corresponds to the single layer graphene, while the bright region below corresponds to the three layer graphene.

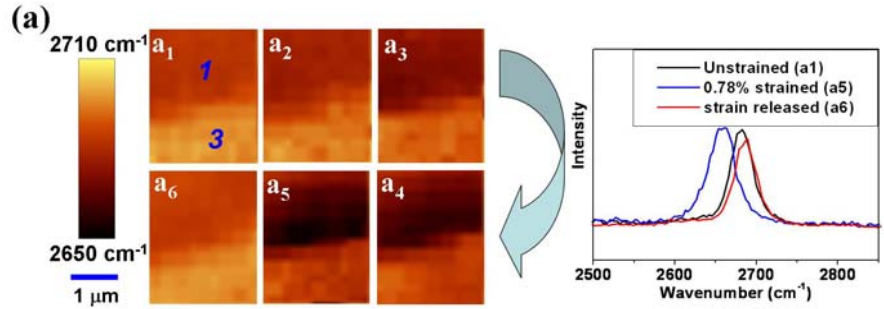

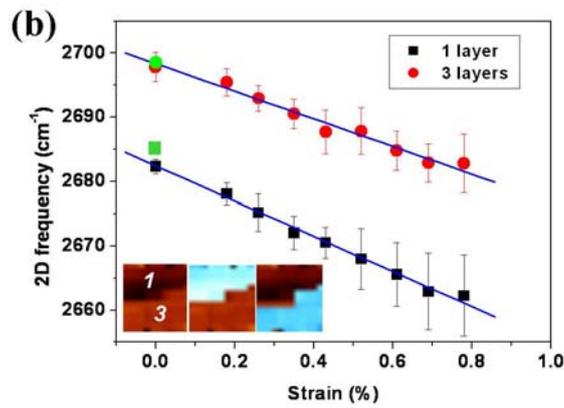

Figure 2 (a) The 2D frequency Raman images of unstrained (a1), 0.18% (a2), 0.35% (a3), 0.61% (a4), and 0.78% (a5) strained, and relaxed (a6) graphene. The scale bar of all the images is 2650 to 2710 cm$^{-1}$. The Raman spectra on the right-hand side are taken from 2D band SLG of a1, a5 and a6. (b) The analyzed 2D band frequency of single (black squares) and three (red circles) layer graphene under different uniaxial strain, from the highlighted area of inset figures. The green square/circle is the frequencies of relaxed graphene. The blue lines are the curve fit to the data. The slope is -27.8 cm$^{-1}$/% for single layer graphene, and -21.9 cm$^{-1}$/% for three layer graphene.

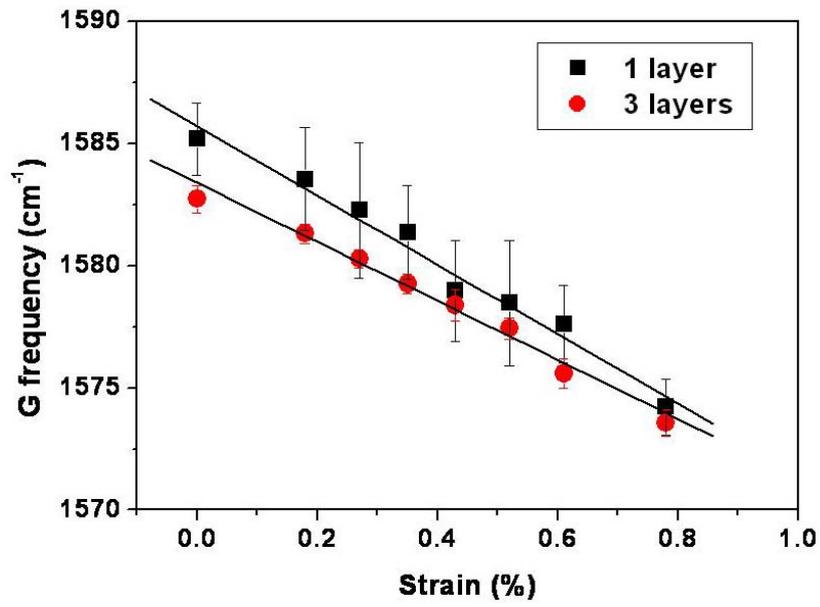

Figure 3. The G band frequency of single layer (black squares) and three layer (red circles) graphene under uniaxial strain. The black lines are the curve fit to the data. The slope is -14.2 cm$^{-1}$/% for single layer graphene, and -12.1 cm$^{-1}$/% for three layer graphene.

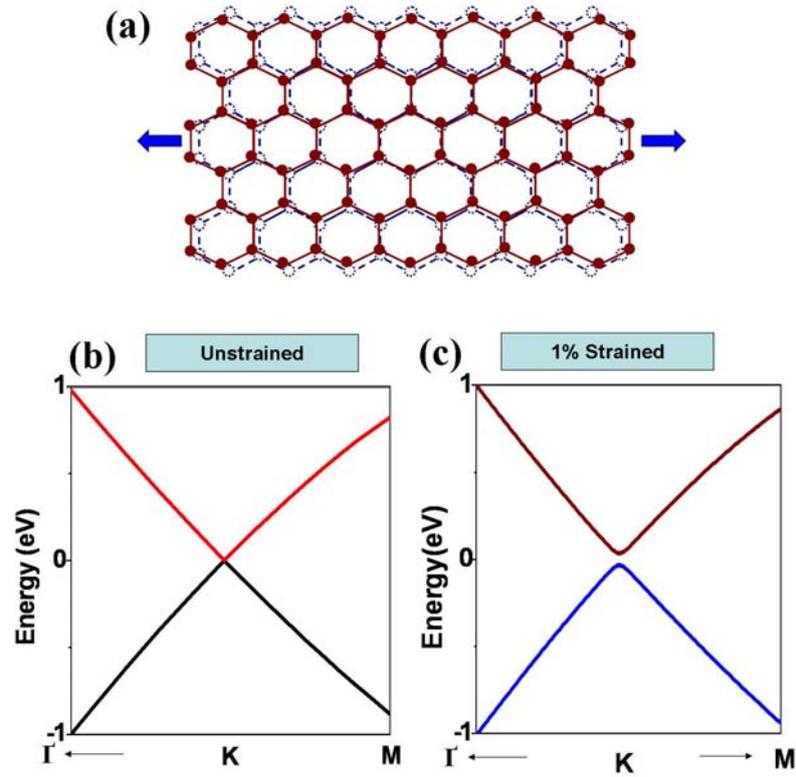

Figure 4. (a) Schematic representation of the effect of uniaxial tensile stress on a graphene supercell. The dashed (solid) lattices indicate the unstrained (strained) graphene. Calculated band structure of unstrained (b) and 1% tensile strained (c) graphene. A band gap is clearly seen on the band structure of strained graphene.

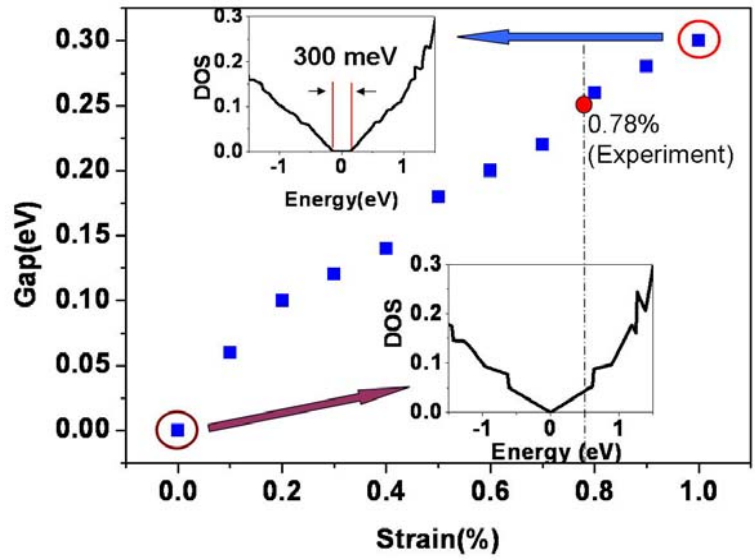

Figure 5. The bandgap of strained graphene with the increase of uniaxial tensile strain on graphene. The magnitude of gap is determined by the gap opening of density of states. The insets show the calculated density of states (DOS) of unstrained and 1% tensile strained graphene. The dash line and red solid dot indicate the calculated bandgap of graphene under the highest strain (0.78%) in our experiment.